\documentclass[reprint,amsmath, amssymb, aps,prb, superscriptaddress, floatfix, footinbib,longbibliography]{revtex4-2}
\usepackage[caption = false]{subfig}
\usepackage{graphicx}
\usepackage{float}
\usepackage{physics}
\usepackage{xcolor}
\usepackage{dcolumn,multirow, tabularx}
\usepackage{lipsum}
\usepackage{diagbox}

\begin{document}

\title{Accurate quantum Monte Carlo forces for machine-learned force fields:\\ Ethanol as a benchmark}
\author{E. Slootman}
\affiliation{Computational Chemical Physics, MESA\textsuperscript{+} Institute for Nanotechnology, University of Twente, PO Box 217, 7500 AE Enschede, The Netherlands}
\author{I. Poltavsky}
\affiliation{Physics and Materials Science Research Unit, University of Luxembourg, L-1511 Luxembourg, Luxembourg}
\author{R. Shinde}
\affiliation{Computational Chemical Physics, MESA\textsuperscript{+} Institute for Nanotechnology, University of Twente, PO Box 217, 7500 AE Enschede, The Netherlands}
\author{J. Cocomello}
\affiliation{Computational Chemical Physics, MESA\textsuperscript{+} Institute for Nanotechnology, University of Twente, PO Box 217, 7500 AE Enschede, The Netherlands}
\author{S. Moroni}
\email{moroni@democritos.it}
\affiliation{CNR-IOM DEMOCRITOS, Istituto Officina dei Materiali, and SISSA Scuola Internazionale Superiore di Studi Avanzati, Via Bonomea 265, I-34136 Trieste, Italy}
\author{A. Tkatchenko}
\email{alexandre.tkatchenko@uni.lu}
\affiliation{Physics and Materials Science Research Unit, University of Luxembourg, L-1511 Luxembourg, Luxembourg}
\author{C. Filippi}
\email{c.filippi@utwente.nl}
\affiliation{Computational Chemical Physics, MESA\textsuperscript{+} Institute for Nanotechnology, University of Twente, PO Box 217, 7500 AE Enschede, The Netherlands}
\date{\today}

\begin{abstract}
 Quantum Monte Carlo (QMC) is a powerful method to calculate accurate energies and forces for molecular systems. In this work, we demonstrate how we can obtain accurate QMC forces for the fluxional ethanol molecule at room temperature by using either multi-determinant Jastrow-Slater wave functions in variational Monte Carlo or just a single determinant in diffusion Monte Carlo. The excellent performance of our protocols is assessed against high-level coupled cluster calculations on a diverse set of representative configurations of the system. Finally, we train machine-learning force fields on the QMC forces and compare them to models trained on coupled cluster reference data, showing that a force field based on the diffusion Monte Carlo forces with a single determinant can faithfully reproduce coupled cluster power spectra in molecular dynamics simulations. 

\end{abstract}

\maketitle

\section{Introduction}

Accurate forces are crucial to perform geometry relaxation and  molecular dynamics (MD) simulations. Classical force fields, which are widely used for such purpose, are often parameterized to reproduce quantum chemical data obtained with approaches such as coupled cluster (CC) or density functional theory (DFT). 
Unfortunately, these force fields cannot always easily capture effects which are fundamentally quantum mechanical. Moreover, their accuracy is intrinsically limited by the predefined functional form, which is in general unknown. For instance, for a system as simple as ethanol at room temperature, MD trajectories based on classical force fields like AMBER \cite{Amber} cannot faithfully explore the potential-energy surface. Consequently, the resulting dynamics does not correctly sample the statistical occupational weights of the hydroxyl rotor group~\cite{Chmiela2018}. 

Machine-learning (ML) force fields \cite{Unke2021} enable performing long MD simulations of \textit{ab initio} quality without the need for expensive quantum chemical calculations at every time step, given a sufficient amount of training data. These ML models are often trained on DFT energies and forces. \cite{Chmiela2017,Chmiela2019,Chmiela2018,Chmiela2023,nequip,orbnet,physnet,schnet,smithANI1ExtensibleNeural2017,so3krates,spookynet,koFourthgenerationHighdimensionalNeural2021,gasteiger_gemnet_2021,GAP,deringerGaussianProcessRegression2021,christensenFCHLRevisitedFaster2020,behlerGeneralizedNeuralNetworkRepresentation2007}
Unfortunately, such a procedure can be unreliable due to the use of approximate functionals as, for instance, whenever additional corrections for DFT must be introduced to capture dispersion interactions.
Then, the accuracy of the DFT reference data must be assessed against highly correlated methods such as coupled cluster (CC) approaches. The most accurate flavors of coupled cluster are however computationally demanding and therefore limited to relatively small molecules.

Quantum Monte Carlo (QMC) calculations can be instrumental in generating the needed reference data for accurate machine-learning potentials. Although QMC is computationally expensive, it provides highly accurate energies and forces, and scales favorably with system size also when forces are computed~\cite{Sorella2010,Filippi2016,Assaraf2017}. Calculating atomic forces in quantum Monte Carlo has been an active field of research and different algorithms and approximations have been put forward for this purpose~\cite{Reynolds1985,Badinski2010,Assaraf2003,Chiesa2005,Filippi2000,Moroni2014,VanRhijn2022,Nakano2023}. 
The use of QMC to construct machine-learning force fields is a relatively new field that has seen applications in the description of high-pressure hydrogen \cite{Tirelli2022,Niu2023,Tenti2023} and in molecular systems \cite{Huang2023, Huang2023a}. Recently, the effect of the statistical noise on the resulting potentials has also been investigated \cite{Ceperley2023}.

Here, we show how QMC can yield forces as accurate as those computed with the ``golden standard'' of quantum chemistry, CCSD(T), over a large set of configurations of the fluxional ethanol molecule at room temperature.
In particular, competitive accuracy can be obtained either in variational Monte Carlo (VMC) using multi-determinant wave functions or in diffusion Monte Carlo (DMC) with the affordable variational-drift-diffusion approximation \cite{Filippi2000,Moroni2014} and just a single determinant. 
Since ethanol is characterized by weak intra-molecular interactions, we also compare our results with DFT calculations treating dispersion interactions with the Tkatchenko-Scheffler (TS)~\cite{Tkatchenko2009} or the many-body dispersion (MBD)~\cite{Tkatchenko2012}  approaches.
Finally, we demonstrate the very good performance of the ML potentials trained on QMC forces using the sGDML model~\cite{Chmiela2019} on unseen test datasets as well as by reproducing the power spectra obtained from MD simulations with CCSD(T) models.

The manuscript is organized as follows. The algorithms to compute the QMC forces and the choice of wave function are described  in Sec.~\ref{sec:method} and the computational details given in Sec.~\ref{sec:computational}. The QMC results and the performance of corresponding ML potentials are discussed in Sec.~\ref{sec:results}. We conclude in Sec.~\ref{sec:conclusion}.

\section{Method}
\label{sec:method}

\subsection{QMC forces}

In QMC \cite{QMC1, QMC2, QMC3}, the energy is computed as
\begin{equation}
    E = \int d \mathbf{R} E_L(\mathbf{R}) P(\mathbf{R})\equiv\langle E_L\rangle_P,
\end{equation}
where $\mathbf{R}$ is the coordinates of the electrons, $E_L(\mathbf{R})={\mathcal{H}\Psi(\mathbf{R})}/{\Psi(\mathbf{R})}$ is the local energy for a given trial wave function $\Psi(\mathbf{R})$, and $P(\mathbf{R})$ is the probability distribution sampled in the QMC run. In VMC, this is equal to $P_\text{VMC}(\mathbf{R}) = |\Psi(\mathbf{R})|^2/\int d\mathbf{R}|\Psi(\mathbf{R})|^2$ and, in DMC, $P_\text{DMC}(\mathbf{R}) = \Phi(\mathbf{R})\Psi(\mathbf{R})/\int d\mathbf{R}\Phi(\mathbf{R})\Psi(\mathbf{R})$ where $\Phi(\mathbf{R})$ is the fixed-node solution. 

The nuclear forces are obtained by taking the derivative of the energy with respect to the nuclear coordinates,
\begin{eqnarray}
    F &=& -\nabla_\alpha E \\
    &=&-\langle\nabla_\alpha E_L(\mathbf{R}) + (E_L(\mathbf{R}) - E)\nabla_\alpha \ln P(\mathbf{R})\rangle_P.\nonumber
    \label{eq:force}
\end{eqnarray}
While the derivative of the distribution function in VMC can be readily performed to compute forces, the distribution function in DMC is not known in closed form but is sampled via a stochastic implementation of the power method through the repeated application of the importance sampled Green function $\mathcal{G}(\mathbf{R}^\prime,\mathbf{R})=\Psi(\mathbf{R}^\prime)\langle\mathbf{R}^\prime| \exp[-\tau(\mathcal{H}-E_\text{T})]|\mathbf{R}\rangle/\Psi(\mathbf{R})$ with $\tau$ the time-step and $E_\text{T}$ an energy shift. Therefore, once equilibrium is reached, $P_\text{DMC}$ is given by 
\begin{eqnarray}
 P_\text{DMC}(\mathbf{R}_n)&=&\int d\mathbf{R}_{n-1}\ldots d\mathbf{R}_{n-k}\\
 &\times& \prod_{i=n-k}^{n-1}\mathcal{G}(\mathbf{R}_{i+1},\mathbf{R}_{i})P_\text{DMC}(\mathbf{R}_{n-k})\,,\nonumber
 \label{P_dmc}
\end{eqnarray}
where $n$ is the last iteration.
The nuclear force in DMC can then be rewritten as
\begin{eqnarray}
 F_\text{DMC}=&-&\langle \nabla_\alpha E_L(\mathbf{R}_n)+[E_L(\mathbf{R}_n)-E]\\
  &\times& \sum_{i=n-k_\text{hist}}^{n-1}\nabla_\alpha\ln {\cal G}(\mathbf{R}_{i+1},\mathbf{R}_{i}) \rangle_{P_\text{DMC}}\,.\nonumber
 \label{force_dmc}
\end{eqnarray}
where $k_\text{hist}$ has to be 
larger than the correlation time 
between $E_L$ and $\nabla_\alpha\ln {\cal G}$ along the random walk~\cite{Moroni2014}.
The importance-sampled Green function must be approximated and, in the limit of small time-steps, becomes 
\begin{equation}
 \mathcal{G}(\mathbf{R}',\mathbf{R})=\frac{e^{-[\mathbf{R}'-\mathbf{R}-V(\mathbf{R})\tau]^2/2\tau}e^{S(\mathbf{R}',\mathbf{R})}}{(2\pi\tau)^{3N/2}},
\end{equation}
where $V(\mathbf{R})=\nabla\Psi(\mathbf{R})/\Psi(\mathbf{R})$ and $S(\mathbf{R}',\mathbf{R})=\tau\{E_{\rm T}-[E_\text{L}({\bf R}')+E_\text{L}({\bf R})]/2\}$. Modified expressions of $V$ and $S$ are used in actual calculations~\cite{Umrigar1993} and the bias due to the short-time approximation can be removed by extrapolating the results at zero time-step. 

While it is possible to compute forces in DMC which are fully compatible with the derivative of the fixed-node DMC energy at any given time-step~\cite{VanRhijn2022}, the derivative of the drift-diffusion part of the Green function introduces larger fluctuations in the force estimator. Therefore, we consider here an estimator of the DMC force in the so-called variational drift-diffusion (VD) approximation~\cite{Filippi2000, Moroni2014}, which only includes the derivative of the branching factor and approximates the derivative of the drift-diffusion contribution by the VMC estimator,
\begin{eqnarray}
	\label{eq:vd}
    F_{\text{VD}} &=& -\langle\nabla_\alpha E_L(\mathbf{R}_n) + [E_L(\mathbf{R}_n) - E]    \\    
&\times&[ \nabla_\alpha P_\text{VMC}(\mathbf{R}_n) +
\sum_{i=n-k_\text{hist}}^{n-1}\nabla_\alpha S(\mathbf{R}_{i+1}, \mathbf{R}_i)]\rangle_{P_{\text{DMC}}}.\nonumber
\end{eqnarray}
Intuitively, this approximation can be derived by 
regarding the random walk in the standard DMC algorithm (drift and diffuse a walker, accept or reject the move, and reweight by the branching factor) as simply reweighting the VMC distribution by the branching factor.  

Computationally, the VD approximation comes at no additional cost since the energy derivatives required for the sum have already been calculated at earlier time-steps. Furthermore, as shown in our calculations, the statistical fluctuations in the VD forces are nearly the same as those obtained when computing the even simpler approximate DMC force introduced by Reynolds \textit{et al.}~(RE)~\cite{Reynolds1985}, which 
computes the VMC force estimator on the DMC distribution,
\begin{eqnarray}
    F_{\text{RE}} &=& -\langle \nabla_\alpha E_L(\mathbf{R})+ [E_L(\mathbf{R}) - E] \\
   &\times&\nabla_\alpha \ln P_{\text{VMC}}(\mathbf{R})\rangle_{P_{\text{DMC}}}\,. \nonumber
\end{eqnarray}
This approximation can be partially corrected by considering the generalized hybrid estimator, $F_{\text{RE-hybrid}}=2F_{\text{RE}}-F_{\text{VMC}}$ at the cost of increased statistical fluctuations \cite{Assaraf2003}.

In general, in addition to the explicit dependence of the energy on the nuclear coordinates through the potential and the trial wave function when an atom-centered basis set is used, there is an implicit dependence through the variational parameters, $p_i$. Consequently, the force acquires an additional term, namely,
\begin{eqnarray}
    F = -\frac{\partial E}{\partial \alpha} - \sum_i \frac{\partial E}{\partial p_i} \frac{\partial p_i}{\partial \alpha}\,,
\end{eqnarray}
where the second term vanishes if the energy is optimal with respect to the parameter variations. Since we fully optimize the wave function in energy minimization at the VMC level, this additional contribution is equal to zero and our VMC forces are fully consistent with the corresponding energy. In DMC, neglecting this term leads in principle to a bias in the corresponding forces, which has however been shown to be quite small if the wave function is fully optimized in VMC, or partially optimized but of sufficient quality like when a multi-determinant expansion is employed~\cite{Moroni2014}.

Finally, all force estimators described above obey a zero-variance principle in the limit of the trail wave function and its derivatives being exact but, for an approximate trial function, display an infinite variance. In VMC, to cure this problem, we employ a guiding wave function which differs from the trial function close to the nodes and is finite at the nodes~\cite{Attaccalite2008}, where we use $d=|\nabla\Psi/\Psi|$ as a measure of the distance from the nodes. While it is possible to adapt this regularization to the computation of DMC forces, 
this has the downside of promoting walkers close to the nodes. Therefore, in the computation of DMC forces, we adopt instead the reguralization scheme from Ref.~\citenum{Pathak2020}, where the force estimator is simply multiplied by a function $f_\epsilon(x)=9 x^2 - 15 x^4 + 7 x^6$ if $x=d/\epsilon < 1$ and $\epsilon$ is chosen sufficiently small to have a negligible bias.

\subsection{Trial wave function}

We employ so-called Jastrow-Slater wave functions of the form
\begin{equation}
    \Psi = \mathcal{J}\sum_i c_i \mathcal{D}_i,
\end{equation}
where $\mathcal{D}_i$ are Slater determinants of single-particle orbitals and $\mathcal{J}$ is the Jastrow correlation factor, which contains electron-electron and electron-nucleus correlation terms~\cite{Jastrow}. All wave function parameters (Jastrow, orbital, and linear coefficients) are fully optimized in energy minimization at the VMC level. 

The determinantal component is here either a single determinant or a multi-determinant expansion generated in an automatic manner with the configuration interaction using a perturbative selection made iteratively (CIPSI) approach~\cite{Huron1973}.  Starting from a wave function expanded on a set of determinants in a given space $S$,
\begin{equation}
    \Psi^{\text{CIPSI}} = \sum_{\mathcal{D}_i\in S}c_i \mathcal{D}_i\,,
\end{equation}
the approach builds expansions by iteratively selecting determinants based on their second-order perturbation (PT2) energy contribution obtained via the Epstein-Nesbet partitioning of the Hamiltonian~\cite{Epstein1926,Nesbet1955},
\begin{eqnarray}
\delta E_{\gamma}^{(2)} = \frac{|\langle\gamma|\mathcal{H}|\Psi^{\rm CIPSI}\rangle|^2}{\langle\Psi^{\rm CIPSI}|\mathcal{H}|\Psi^{\rm CIPSI}\rangle - \langle\gamma|\mathcal{H}|\gamma\rangle}\,,
\label{eq:en-pt2}
\end{eqnarray}
where $|\gamma\rangle$ denotes a  determinant outside the current CI space that is connected to $S$ by $\mathcal{H}$.  The total PT2 energy contribution, $E^{\text{(PT2)}}$, goes to zero as the expansion approaches the full CI (FCI) limit. 

We are here interested in computing forces on different structural configurations and want to achieve a balanced CIPSI description of the determinantal component of the QMC wave function across the ground-state potential energy surface of ethanol. As a
measure of the quality of a CIPSI wave function, we use 
its PT2 energy contribution, which represents an
approximate estimate of the error of the expansion with respect to FCI. Therefore, given the chosen expansion and its energy PT2 correction for an arbitrary reference configuration, we generate expansions for the other configurations by
  matching the reference $E^{\text{(PT2)}}$. In general, the procedure will result in expansions of different length at the different geometries.

\section{Computational Details}
\label{sec:computational}

The QMC calculations are carried out with the CHAMP code~\cite{CHAMP}. We employ scalar-relativistic energy consistent Hartree-Fock pseudopotentials and the correlated-consistent Gaussian basis sets specifically constructed for these pseudopotentials~\cite{Burkatzki2007,BFD_H2013}. For most calculations, we use the cc-pVTZ basis set and perform convergence tests with the cc-pVQZ basis set. As shown in Table S1 for a representative configuration and a single-determinant wave function, the use of a cc-pVTZ basis yields VMC forces which are converged with respect to the basis set.

All wave function parameters (Jastrow, orbital, and CI coefficients) are optimized by minimizing the energy in VMC using the stochastic reconfiguration method~\cite{Sorella2007} in a low-memory implementation~\cite{Neuscamman2012}. To cure the diverging variance of the force estimator, we employ a 
node cutoff parameter $\epsilon$ of 0.1 a.u.\ in VMC and 0.05 a.u.\ in DMC. 
In the DMC calculations, we treat the pseudopotentials beyond the locality approximation using the T-move algorithm~\cite{Casula2006} and employ a time-step of 0.005 a.u.\ which ensures converged VD forces as shown in Fig.~S5. A value of 900 is used for $k_\text{hist}$ (Eq.~\eqref{eq:vd}) and the dependence of the VD forces on this parameter is illustrated in Fig.~S6. In the regularization procedure~\cite{Pathak2020}, our choice of 0.05 a.u.\ for $\epsilon$ yields a negligible bias compared to the statistical error as shown in Section S3.  

We perform the HF calculations with the program GAMESS(US)~\cite{Schmidt1993} and generate the CIPSI wave functions with Quantum Package~\cite{Garniron2019} using the same pseudopotential and basis sets as in QMC. The interface of both these codes with CHAMP uses the TREXIO library~\cite{Posenitskiy2023}. The Psi4 package~\cite{Smith2020} is employed for the all-electron coupled cluster calculations with Dunning's correlated consistent basis sets~\cite{Dunning1989}.

The machine learning models for ethanol use 100 training and 100 validation configurations (thereafter referred as set $A$), 
obtained by clustering a set of 2000 representative configurations (set $B$) down to 200 (set $A$), based on their geometry and energy. To this aim, we first split the 2000 configurations into 40 clusters based on their geometry, using the agglomerative clustering algorithm and, then, split each cluster into 5 clusters based on energy using the k-means method. Afterward, configurations closest to the centroids of each cluster are selected to form the training set (see Ref.~\citenum{Fonseca2021} for more details). 
The initial 2000 configurations (set $B$) were extracted from a long MD trajectory based on DFT calculations with the PBE-TS functional~\cite{Chmiela2017} by sampling them according to the energy distribution in this trajectory.
For this purpose, we use the implementation from the symmetric gradient domain machine learning (sGDML) software package~\cite{Chmiela2018}: the energies of the configurations are histogrammed and a number of configurations proportional to the height of the histogram is then selected randomly within each bin.
A second set of 2000 configurations (set $C$) is clustered from the complete MD trajectory according to the procedure followed for set $A$.
Therefore, in contrast to set $B$, set $C$ equally represents different possible molecular geometries and energy states irrespective of their statistical probability in the reference dataset.
 
The sGDML models are trained on set $A$ with energy and forces computed with different \textit{ab initio} methods, namely, QMC (i.e.\ VMC, RE, RE-hybrid, and VD DMC),  CCSD(T) with the cc-pVTZ and cc-pVQZ basis sets, and DFT PBE-TS and PBE0-MBD. For the DFT and CCSD(T)/cc-pVTZ, sGDML models are also trained on the larger set B.  The error of the obtained force fields is analyzed using the open-source FFAST software package~\cite{Fonseca2023}. 

The classical MD simulations are carried out with a time-step of 0.2~fs at a temperature of 300~K employing a Langevin thermostat with a time constant of 100~fs, using the i-PI package~\cite{Kapil2019}. The total duration of the MD trajectories is 0.6~ns.

\section{Results and discussion}
\label{sec:results}

We demonstrate the performance of QMC forces on the fluxional ethanol molecule, characterized by intra-molecular dispersion forces between the the hydroxyl and methyl rotors, namely, between the lone pairs of the oxygen and the partially positive charges of the hydrogen atoms. 
We compute here the QMC forces with different algorithms and wave functions, and discuss the impact of these choices on the corresponding ML models constructed with the sGDML framework for which ethanol is particularly suitable given its many symmetries. We also compare the QMC results to those obtained with two DFT functionals, namely, PBE-TS and PBE0-MBD. 

As reference, we calculate the CCSD(T) forces also with a cc-pVQZ basis on set $A$, while, on the larger set $B$, we only perform the CCSD(T) calculations with the smaller cc-pVTZ basis set. We discuss the basis set convergence of the CC results below and in Section S2.

\subsection{Quality of the forces}

\begin{figure}[thb]
    \centering
    \includegraphics[]{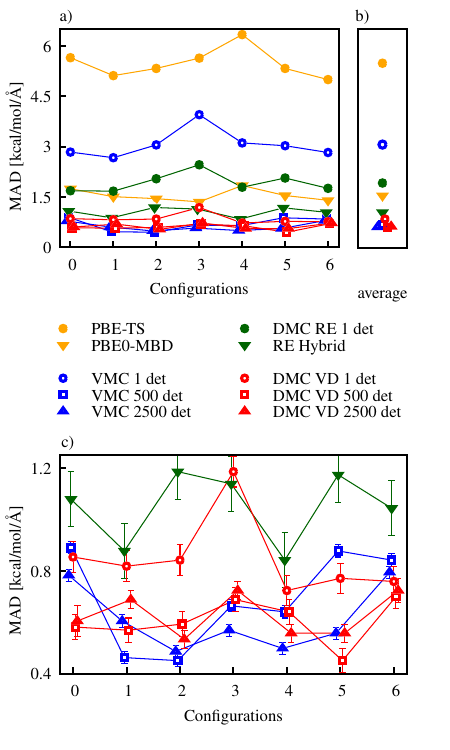}
    \caption{a) Mean absolute deviation (MAD, kcal/mol/\AA) of the forces computed with different methods, compared to CCSD(T)/cc-pVQZ for 7 representative configurations of room-temperature ethanol; b) average MAD for each method; c) zoomed-in version of the MADs 
    with their statistical errors.}
    \label{fig:lines}
\end{figure}

We begin our investigation by analyzing in detail the behavior of the various methods on seven representative configurations (selected to represent a variety of molecular geometries and energies within subset $B$) of ethanol at room temperature.  
In Fig.~\ref{fig:lines}, we show the mean absolute deviation (MAD) of the forces with respect to CCSD(T)/cc-pVQZ for each configuration as well as the average MAD over the seven configurations. Of the methods investigated here, PBE-TS is the least accurate, while the use of PBE0-MBD yields significantly higher accuracy for this system.  Moreover, PBE0-MBD demonstrates a relatively small dependence of the MAD upon the specific configuration.  

VMC forces with a one-determinant wave function display a significant error that lies between the two DFT methods. Using the CIPSI procedure to go beyond a single determinant, we construct two expansions for each configuration, matching two different values of the total PT2 energy correction to ensure a consistent quality of the wave function across different geometries. More specifically, for configuration 2, we generate two expansions of about 500 and 2500 determinants, yielding a PT2 correction of $-0.676$ and $-0.639$ a.u., respectively, and use these two energy values as target in the CIPSI generation at the other configurations. The number of determinants in the resulting expansions ranges between 309-995 and 2098-3484, respectively. Further information on the convergence of the QMC results as a function of determinantal number is given in Section S5. 

The results obtained with the CIPSI-based fully-optimized Jastrow-Slater wave functions are shown in Fig.~\ref{fig:lines} and denoted for simplicity as ``500 det'' and ``2500 det''. At the VMC level, the 500-det wave function yields a big improvement on the one-determinant forces, surpassing the PBE0-MBD results. Further enlargening the expansion with the use of the 2500-det wave functions improves only marginally the accuracy. The relative flatness of the 500-det and 2500-det VMC lines for different geometries is a clear indication of the success of the PT2-matching construction in yielding determinantal expansions of comparable quality when employed in a Jastrow-Slater wave function. 

When carrying out DMC calculations on these VMC-optimized wave functions, we find that the VD forces perform very well already in the one-determinant case. On the contrary, the RE forces show some improvement over VMC but do not beat the accuracy of DFT/PBE0-MBD. Correcting these forces via the RE-hybrid estimator brings the forces close to the VD ones at the expense of larger statistical fluctuations (see Fig.~\ref{fig:lines}c). The use of DMC-VD in combination with the multi-determinant wave functions shows in general no further, significant improvement compared to the one-determinant VD case: The VMC and DMC-VD forces for the multi-determinant wave functions and the DMC-VD forces for the one-determinant wave function, have roughly the same MAD with respect to CCSD(T).  

The one-determinant DMC-VD case for configuration 3 is clearly an outlier, displaying a larger deviation from the reference. This can be explained by inspecting the geometry of the molecular configuration (shown in Fig.~S1), which is quite distorted with an angle of the methyl group characteristic of a region near a barrier in the potential energy surface. The wave function of such a configuration has therefore a more correlated character and must include multiple determinants to be accurately described. In fact, the MAD in DMC-VD for configuration 3 reduces when enlargening the expansion from one to 995 and further to 3484 determinants. 
he accuracy of the QMC calculations has been pushed to a level where the remaining discrepancy of the forces with respect to the all-electron CCSD(T)/cc-pVQZ results can be attributed to the use of pseudopotentials in QMC, and/or to residual basis set errors in CCSD(T), as further elaborated in Section S2.  

\begin{figure}[thb]
    \centering
    \includegraphics[]{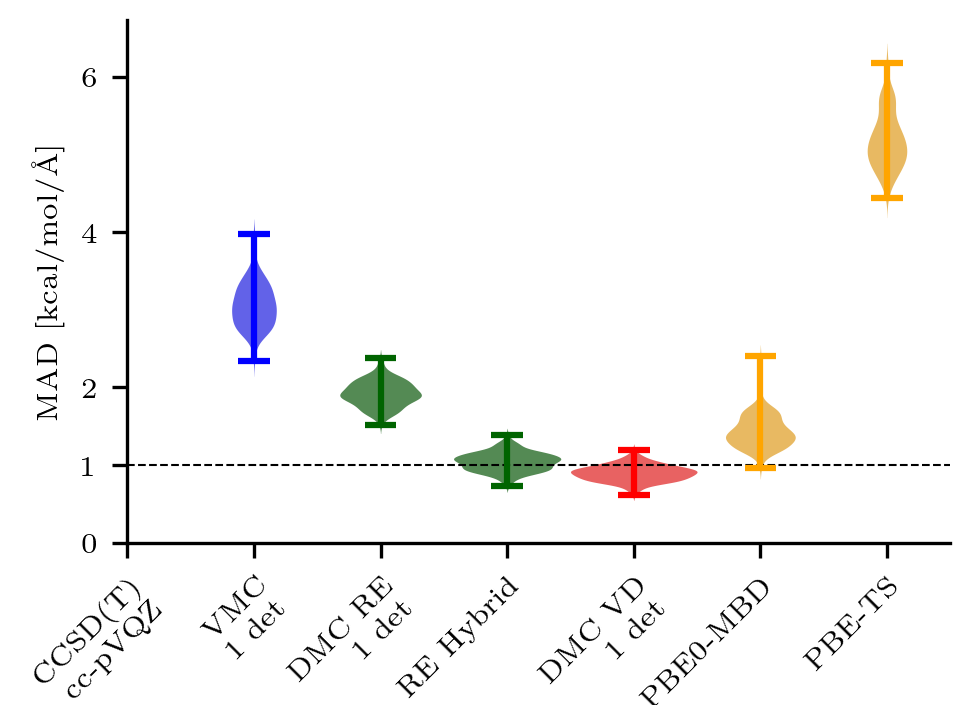}
    \caption{Mean absolute deviation (kcal/mol/\AA) of the QMC and DFT forces with respect to CCSD(T)/cc-pVQZ for 200 configurations (set A) of room-temperature ethanol. }
    \label{fig:violin}
\end{figure}

To verify the robustness of these findings over a larger dataset, VMC and DMC calculations are performed with a one-determinant fully-optimized Jastrow-Slater wave function on 200 representative configurations (set $A$) of room-temperature ethanol. The quality of the QMC results is again assessed against CCSD(T)/cc-pVQZ and also compared to the outcome of the DFT calculations. The MADs of all configurations with respect to coupled cluster are plotted in Fig.~\ref{fig:violin} and the average MAD values reported in Table~\ref{tab:200res}. 

The results show the same pattern as observed for the 7 configurations of Fig.~\ref{fig:lines}, corroborating the findings above. In particular, VD-DMC lowers the errors and their spread compared to the VMC and the RE forces, and shows once again that, for this system, the one-determinant DMC VD forces are very accurate. The use of RE-hybrid offer a relatively large improvement on the RE forces.  However, since it comes with a statistical error more than twice as large, there is no real use case for this method.   

\begin{table}[thb]
    \begin{tabular}{l|l}
        Method & MAD\\
        \hline
        VMC 1 det    & 3.055(2)\\
        DMC RE 1 det & 1.920(4)\\
        RE-hybrid    & 1.046(8)\\
        DMC VD 1 det & 0.899(4)\\
        PBE-TS       & 5.181\\
        PBE0-MBD      & 1.431\\
    \end{tabular}
    \caption{Average mean absolute deviation (kcal/mol/\AA) of the different methods with respect to CCSD(T)/cc-pVQZ over 200 configurations (set $A$) of ethanol. The statistical error on the last digit is indicated in brackets.}
    \label{tab:200res}
\end{table}

\begin{figure*}[thb]
    \centering
    \includegraphics[]{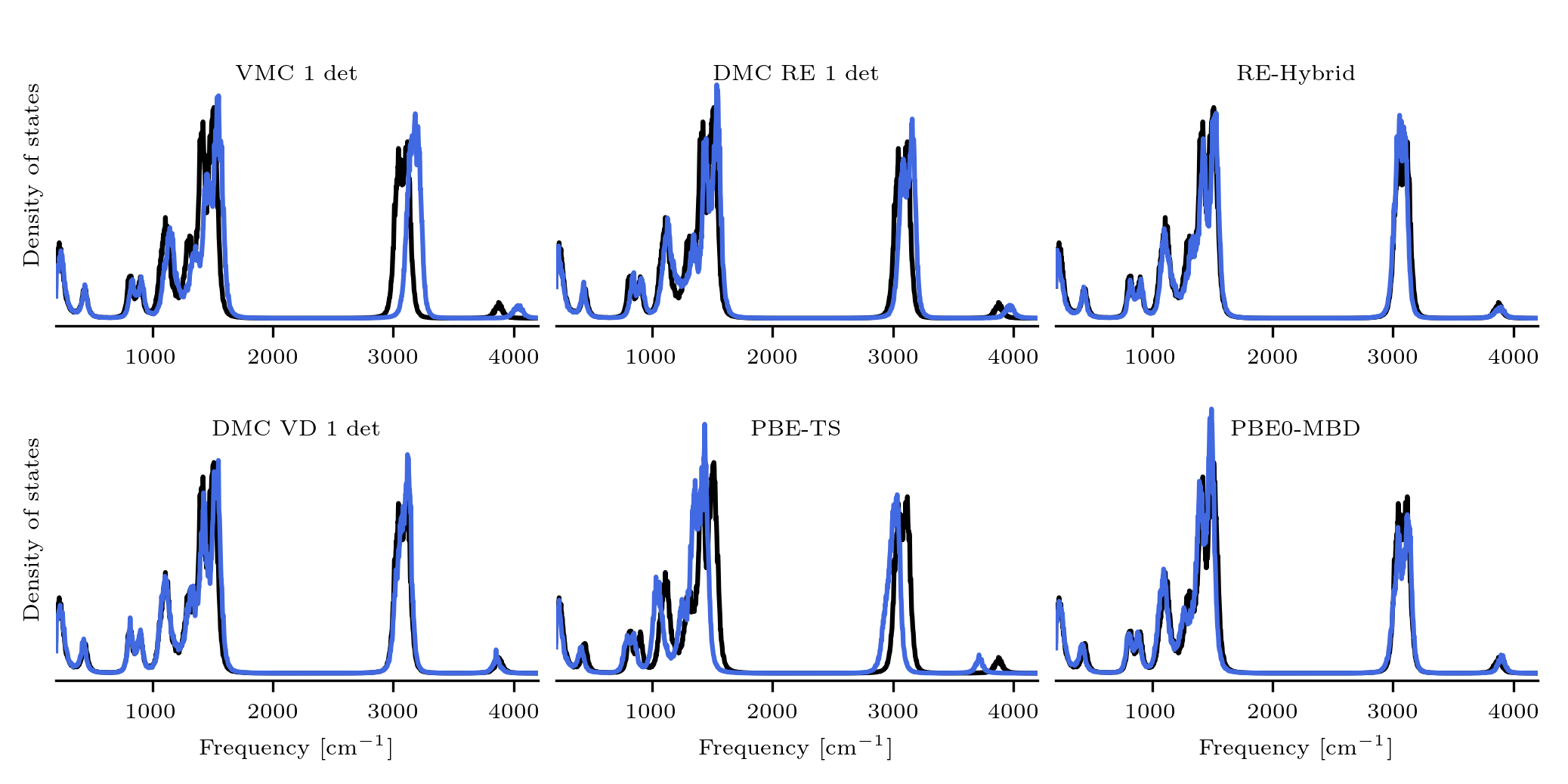}
    \caption{Vibrational spectra of ethanol at room temperature computed with the various ML models (blue) compared to the CCSD(T)/cc-pVQZ (black) vibrational spectrum.}
    \label{fig:spectrum}
\end{figure*}

\subsection{Effect on Machine-Learning Force Fields}

With the forces computed with the different methods on the 200 configurations of set $A$ (Fig.~\ref{fig:violin}), we generate ML force fields using the sGDML model, using half of the data as training and the other half as validation points. 
For CCSD(T)/cc-pVTZ, PBE-TS, and PBE0-MBD, we show in Section S6 that constructing the ML force fields on a larger set of configurations does not affect the relative quality of the models. 

\begin{table}[tbh]
    \begin{tabular}{l|c|c|c|c}
    \diagbox{model}{dataset} & \begin{tabular}[c]{@{}c@{}}$A \subset B$\\200 \\Q\end{tabular} & \begin{tabular}[c]{@{}c@{}}$A \subset B$\\ 200 \\T\end{tabular} & \begin{tabular}[c]{@{}c@{}}$B$\\ 2000\\T \end{tabular} & \begin{tabular}[c]{@{}c@{}}$C$\\ 2000\\T \end{tabular} \\ \hline
    VMC       & 3.2 & 3.4 & 3.4 & 3.4 \\
    RE        & 2.2 & 2.4 & 2.5 & 2.6 \\
    RE Hybrid & 1.5 & 1.8 & 2.0 & 2.2 \\
    VD        & 1.2 & 1.4 & 1.6 & 1.8 \\
    PBE-TS    & 5.3 & 5.3 & 5.3 & 5.3 \\
    PBE0-MBD  & 1.7 & 1.9 & 2.0 & 2.1 \\
    CCSD(T)/cc-pVTZ & 1.0 & 0.7 & 1.2 & 1.4 \\
    CCSD(T)/cc-pVQZ & 0.7 & 1.0 & 1.4 & 1.5 \\
    \end{tabular}
    \caption{Mean absolute deviation (kcal/mol/\AA) of the forces obtained from the ML models on different datasets (A, B, C) against CCSD(T)/cc-pVXZ forces with X=T, Q. These values are calculated with the FFAST software~\cite{Fonseca2023}.}
    \label{tab:dataset}
\end{table}

The performance of the ML models is assessed on three sets of configurations ($A\subset B$, $B$, and $C$) by computing the MAD of the ML forces with respect to the CCSD(T) values calculated with either the cc-pVTZ (sets $A$, $B$, and $C$) or the larger cc-pVQZ basis set (set $A$).  Using coupled cluster with the smaller basis as reference on set $A$ leaves the ordering of the MADs unchanged, justifying the use of cc-pVTZ to evaluate the CCSD(T) reference on the larger $B$ and $C$ sets as shown in Table~\ref{tab:dataset}. For all datasets, we find that the quality of the ML models nicely follows the quality of the underlying {\it ab initio} forces as depicted in Fig.~\ref{fig:violin}. 
Not surprisingly, CCSD(T) displays the smallest MAD since the reference values are computed using the same method.  Note that the mean absolute errors of the ML models on the validation sets are about 1.2--1.3 kcal/mol/\AA\ (see Table S5). A difference of the same magnitude between the force field predictions and the reference data is therefore not significant for practical applications. 

Importantly, we test the ML models on a dataset of 2000 configuration (set $C$), which is totally independent of the datasets ($A$ and, in Section S6,  $B$) used to generate the force fields. This test further confirms that the relative performance of the ML models follows the accuracy of the {\it ab initio} forces. Also on this dataset, we find that the model based on DMC-VD forces yields a smaller MAD than the ones constructed with VMC, other DMC approximations, and DFT.

Finally, to further analyze the behavior of the force field models, we compute the vibrational spectra from the velocity autocorrelation functions in classical MD simulations at room temperature. These can lead the system to regions of the potential energy surface which are not well sampled in the testing datasets. The spectra are shown in Fig.~\ref{fig:spectrum} and compared to the one obtained with the model trained on CCSD(T)/cc-pVQZ.  As regards QMC, we observe again a gradual increase of accuracy moving from VMC, to DMC-RE, and, finally, to DMC VD. This is clearly visible in the overall shift of the spectrum and, in particular, of the C-H vibrational peaks around 3000 cm$^{-1}$, which are clearly overestimated by the VMC model. We note that also DMC-Hybrid and PBE0-MBD perform rather similarly to DMC-VD, while PBE-TS model underestimates the vibrational frequencies.

\section{Conclusion}
\label{sec:conclusion}

We have investigated the use of different algorithms and wave functions for the calculation of forces in QMC for ethanol at room temperature. For this system, a multi-determinant wave function in VMC is found to yield forces of comparable quality to those obtained with a single-determinant wave function and the DMC-VD approach. In both cases, the forces are in excellent agreement with the CCSD(T) values on a representative set of configurations. Employing the generalized hybrid estimator of the RE-Hybrid method also leads to accurate forces  but is of less practical use due to the larger statistical error. Finally, we demonstrated the ability to train accurate machine-learning force fields using QMC. In particular, the sGDML model trained on single-determinant DMC-VD forces is shown to faithfully reproduce the vibrational spectrum of ethanol at room temperature obtained in molecular dynamics simulations with the CCSD(T)-based model. These findings unveil the potential that QMC methods offer in providing forces as reference data for machine-learning force fields, being as accurate as coupled cluster calculations and yet computationally applicable to large molecular systems.

\acknowledgements

E.S., R.S., S.M. and C.F. acknowledge partial support by the European Centre of
Excellence in Exascale Computing TREX --- Targeting Real Chemical
Accuracy at the Exascale. This project has received funding in part from the
European Union's Horizon 2020 --- Research and Innovation program ---
under grant agreement no.~952165.
I.P. and A.T. acknowledge funding in whole, or in part, by the Luxembourg National Research Fund (FNR), grants reference C19/MS/13718694/QML-FLEX and INTER/MERA22/16521502/PHANTASTIC.
The calculations were carried out on the Dutch national supercomputer Snellius with the support of SURF Cooperative.

\bibliography{main.bib}

\begin{thebibliography}{62}
\bibitem [{Jas()}]{Jastrow}%
  \BibitemOpen
  \href@noop {} {}\bibinfo {note} {As Jastrow factor, we use the exponential of the sum of two fifth-order polynomials of the electron-nuclear and the electron-electron distances, respectively, and rescale the inter-particle distances as $R=(1-\exp(-\kappa r))/\kappa$ with $\kappa$ set to 0.6 a.u. We employ different electron-nucleus Jastrow factors to describe the correlation of an elecron with C, O, and H. The total number of free parameters to be optimized in the Jastrow factor is 17 for the systems considered here.}\BibitemShut {Stop}%
\end{thebibliography}

\end{document}